\documentclass[12pt]{article}
\usepackage{graphicx}
\usepackage[breaklinks]{hyperref}
\usepackage{booktabs}
\usepackage{amsmath}
\usepackage{amssymb}

\newtheorem{theorem}{Theorem}[section]
\newtheorem{lemma}[theorem]{Lemma}

\newcommand{\qed}{\nobreak \ifvmode \relax \else
      \ifdim\lastskip<1.5em \hskip-\lastskip
      \hskip1.5em plus0em minus0.5em \fi \nobreak
      \vrule height0.75em width0.5em depth0.25em\fi}

 \def\tads{ \widetilde{ \text{AdS}} }
\begin{document}
\title{AdS spacetimes and isometric embeddings}

%% \author{Your Name}
%% \address{Affiliation, Address}
%% \ead{your@email.address}

\author{Steven Willison\footnote{steven.willison-at-ist.utl.pt}\\
CENTRA, Departamento de Fiısica,
Instituto Superior T´ecnico — IST,\\ Universidade T´ecnica de Lisboa - UTL,\\ Av.
Rovisco Pais 1, 1049-001 Lisboa, Portugal}

%\date{September 22, 2012}

\maketitle

\begin{abstract}
An algebraic global isometric embedding of the nonrotating BTZ
black hole is presented. The ambient spacetime is
$\mathbb{M}^{2,3}$, the 3+2 dimensional flat spacetime. We also
present the analogous embedding for the Euclidean BTZ spacetime
and by performing a kind of double analytic continuation construct
a 1-parameter family of embeddings of cosmological AdS spacetime
into $\mathbb{M}^{2,3}$ which coincide asymptotically with the
embedded BTZ manifold of the appropriate mass. Finally we note
that the family of embeddings of cosmological AdS$_{n}$ into
$\mathbb{M}^{2,n}$ generalises to higher dimensions.
\end{abstract}

{\small (This article is to appear in the proceedings of the
conference: ``Relativity and Gravitation, 100 Years after Einstein
in Prague," June 25 - 29, 2012, Prague, Czech Republic. We give a
brief review of the results obtained in [arXiv:1011.3883[gr-qc]].
Lemma 2.2, is a new result.)}

\section{Some Differential geometric preliminaries}

Anti de Sitter space $\tads$ is a Lorentzian manifold and as such all of the geometry can be understood in terms of intrinsically defined properties. However, to manifest more symmetries one often considers the following model: Let $M^{p,n}$ denote flat spacetime of signature $-p+n$. The submanifold:
\begin{gather}
 \mathcal{A}:= \{X^\mu \in \mathbb{M}^{2,n-1}|  X^\mu X^\nu \eta^{(2,n-1)}_{\mu\nu} =-1\},
\end{gather}
has the same intrinsic geometry as Anti de Stter space $\tads^n$.
By this we mean that the submanifold so defined is diffeomorphic to $S\times\mathbb{R}^3$ and the induced metric is that of a maximally symmetric Lorentzian spacetime with sectional curvatures $-1$.
By representing the spacetime as a submanifold certain geometrical facts become clear:
%Every timelike straight line through the origin intersects two unique points in $\mathcal{H}$;
The intersections with hyperplanes $\mathcal{P}$ through the origin are geodesics, specifically
$\mathcal{A} \cap \mathcal{P}^{2,n-2}$ is a timelike geodesic and $\mathcal{A} \cap \mathcal{P}^{1,n-1}$ is a pair of disconnected spacelike geodesics;
%Similarly, totally geodesic $p$-surfaces are of the form $\mathcal{A} \cap \mathcal{P}^{1,p}$ or  $\mathcal{A} \cap %\mathcal{P}^{2,p-1}$;
$\mathcal{A}$ contains closed timelike curves; the global isometry group is $SO(n,1)$.
This embedding of $\tads^n$ is well known. Two spacetimes of special interest which are locally isometric to $\tads$ are the cosmological AdS spacetime and the BTZ\cite{BTZ} spacetime in 2+1 dimensions. The former is the universal covering space obtained by unwrapping the timelike geodesics. The latter is a quotient space. Therefore the local physics can be described as a constrained theory in flat spacetime using $\mathcal{A}$ described above. However, to adress global issues it can be helpful to have a simple global isometric embedding. Below we present such embeddings for the norotating BTZ spacetime (obtained in \cite{Willison:2010nj}) and also for AdS$^n$.

A smooth \emph{embedding} $\phi: M \to N$ is an injective map such that $\phi(M)$ is homeomorphic to $M$ and $\phi_*$  is injective.
Note $dim (N)\geq dim (M)$.
%In terms of coordinates $(x^\mu)$ on $M$ and $(y^A)$ on $N$, we write $\phi: x \to y(x)$ and $\phi_*: v^\mu \to v^\mu \partial %y^A/\partial x^\mu$.
Let $(M,h)$ and $(N,g)$ be pseudo-Riemannian manifolds.
A smooth embedding $\phi: M\to N$ is an \emph{isometric embedding} if $\phi^* g = h$. If $M$ is the entire maximally extended spacetime we call this a \emph{global isometric embedding}.

\section{Embedding the BTZ and cosmological AdS spacetimes}

In 2+1 dimensional gravity with negative cosmological constant, the Einstein equation in vacuum is equivalent to
$ R^{\mu\nu}_{\quad\kappa \lambda} = -\frac{1}{l^2} (\delta^{\mu}_{\kappa}\delta^\nu_\lambda - \delta^\nu_\kappa  \delta^\mu_\lambda)$.
We set $l =1$.
The spherically symmetric solution of mass $a^2$ has the static form \cite{BTZ}
\begin{gather*}
ds^2 = (r^2 - a^2) d\tau^2 + \frac{dr^2}{r^2-a^2}+r^2 d\phi^2
\end{gather*}
outside of the event horizon ($r =a$). Since we are interested in global embeddings, we introduce the
Kruskal type coordinate system:
\begin{gather*}
 ds^2 = 4\frac{-dt^2 +dx^2 }{(1+t^2-x^2 )^2}
 + a^2\frac{(1-t^2 +x^2)^2}{(1+t^2 -x^2)^2}
 d\phi^2\, .
\end{gather*}
The domain of the coordinates is $-1 <-t^2 +x^2< 1$, $\phi\sim \phi +2\pi$. This covers the maximally extended space-time. The event horizons and bifurcation surface are $x = \pm t$ and $x=t=0$ respectively. Singularities ($t^2 -x^2 =1$) and conformal infinity ($x^2 -t^2 =1$) are not considered part of the spacetime for our purposes.

Let $\mathbb{M}^{2,3}$ denote the flat space with metric $g =
\eta^{(2,3)}_{\mu\nu}dX^\mu dX^\nu:= -(dX^0)^2 + (dX^1)^2 +
(dX^2)^2 + (dX^3)^2 - (dX^4)^2$.
\begin{lemma} \cite{Willison:2010nj}
The nonrotating BTZ black hole spacetime can be globally
isometrically embedded into the region $X^0>0$ of
$\mathbb{M}^{2,3}$. The image is the intersection of quadric
hypersurfaces:
\begin{gather*}  (X^1)^2  +(X^2)^2= \frac{a^2}{1+a^2} (X^0)^2\,  , \quad (X^3)^2- (X^4)^2 = -1 + \frac{1}{1+a^2} (X^0)^2\end{gather*}
The past and future singularities are located at the intersection of the two constraint surfaces with the hyperplane $X^0=0$.
\end{lemma}

%Proof: It can be verified that the following is an embedding
%\begin{align*}
% X^0 & =\sqrt{1+a^2}\left(\frac{1-t^2+x^2 }{1+t^2-x^2}\right)\, ,\\
%X^1 & = a \left(\frac{1-t^2+x^2}{1+t^2-x^2}\right)\cos \phi\,,\
%X^2  = a \left(\frac{1-t^2+x^2}{1+t^2-x^2}\right)\sin \phi\, , \\
%X^3 &= \frac{2x}{1+t^2-x^2}\, ,\
%X^4 = \frac{2t}{1+t^2-x^2}\, .
%\end{align*}
The proof was given in Ref.  \cite{Willison:2010nj}. Combining the
constraint equations we have $X^\mu X^\nu \eta^{(2,3)}_{\mu\nu} =
-1$ therefore a global embedding into $\widetilde{\text{AdS}}_4$
exists. By lifting the restriction $X^0>0$  we obtain two copies
of BTZ joined at the singularity, but it is not a true embedding
at $X^0 = 0$: the tangent space map is not injective (the central
singularity is a conical singularity).

We may make an analytic continuation $X^4 \to iX^4$, whence we obtain an embedding of the Euclidean black hole into $\mathbb{ M}^{(1,4)}$ \footnote{This belongs to a class of immersions of $H^3$ into $H^4$ obtained in Ref. \cite{Nomizu}}.  In fact there is another embedding, of a Euclidean black hole with mass parameter $1/a$, which has the same asymptotic form for large $X_0$. They are related by $(X^1,X^2)\leftrightarrow (X^3 , X^4)$. We shall refer to this as
``thermal AdS."  The reason for this apparently arbitrary distinction is that upon making the analytic continuation $X^4 \to -iX^4$ we obtain now a global embedding of (two copies of) the cosmological AdS$_3$ which coincide asymptotically with the exterior regions of the black hole. All of this is summarised in table \ref{quadratic_table}.
In the case of the embedding of AdS$_3$, the parameter $a$ has no intrinsic geometrical meaning, and therefore no direct physical meaning. We call $2\pi a$ the \emph{extrinsic temperature}\footnote{Another kind of extrinsic notion of temperature, based on  a local embedding modeled on $\mathcal A^3$, was introduced in Ref \cite{Deser:1998xb}.}  in this context since it is the temperature of the embedded BTZ spacetime to which it is asymptotic.

\begin{table}[h]
\caption{\label{quadratic_table}Various embeddings relevant to three dimensional gravity. The left and right colums are related by $X_4\leftrightarrow iX_4$.
%The EBTZ and ``Thermal AdS" are related by $(X_1,X_2)\leftrightarrow (X_3 , X_4)$.
} % title of Table
\centering % used for centering table
\begin{tabular}{ c|c  } % centered columns (2 columns)
\hline
Embedding in $\mathbb{M}_{3,2}$ &   Embedding in $\mathbb{M}_{4,1}$ \\ [0.5ex] % inserts table heading
\hline
BTZ (mass = $a^2$)  & Euclidean BTZ (mass $=a^2$)\\
$( X^1)^2 + (X^2)^2  =\frac{a^2}{1+a^2} (X^0)^2 $, &  $(X^1)^2 +( X^2)^2= \frac{1}{a^2 +1} (X^0)^2 $,
\\
$(X^3)^2 -(X^4)^2  =  \frac{1}{a^2 +1} (X^0)^2 -1 $,& $ (X^3)^2  +(X^4)^2 = \frac{a^2}{1+a^2} (X^0)^2 -1 $,\\
$X^0>0$ ($X^0 = 0$  singular). &$X^0>0$.\\\hline
AdS$_3$  &  ``Thermal AdS$_3$" (mass $= 1/a^2$)\\
$ (X^1)^2 + (X^2)^2  =\frac{a^2}{1+a^2} (X^0)^2 -1$, &  $(X^1)^2 + (X^2)^2= \frac{1}{a^2 +1} (X^0)^2 -1$,\ \
\\
$(X^3)^2 -(X^4)^2  =  \frac{1}{a^2 +1} (X^0)^2  $, & $ (X^3)^2  +(X^4)^2 = \frac{a^2}{1+a^2} (X^0)^2  $,\\
$X^0>0$ (2 copies of AdS).& $X^0>0$. \\
 [1ex] % [1ex] adds vertical space
\hline %inserts single line
\end{tabular}
\end{table}

The complete picture contained in table \ref{quadratic_table} is peculiar to three dimensions and depends on the fact that the black hole and AdS are related by a double Wick rotation which exchanges the role of the angular coordinate with that of the Euclidean time.
However, we are able to present here the following result pertaining to higher dimensions:

\begin{lemma}\label{AdSn_lemma}
 Let $\mathbb{M}_{n,2}$ be pseudo-Euclidean space with two time directions and standard coordinates $X^A = (T, X^1,...,X^n,S)$ and $\alpha$ be a positive real number.
Then the submanifold $\{X^A\in \mathbb{M}_{n,2}| X^AX^B\eta_{AB} =-1;\  (X^n)^2 = S^2 +\frac{\alpha}{1+\alpha^2} T^2;\ T, X^n >0\}$ is homeomorphic to $\mathbb{R}^n$ and globally isometric to AdS$_n$.
\end{lemma}

Proof: We introduce angular coordinates $(\theta^i)$, $i = 1, \dots, {n-2}$ on the unit sphere $\sigma^a\sigma^a =1$ in $\mathbb{R}^{n-1}$. We then consider a cylindrical polar system of coordinates $(\tau,r, \theta^i)$ on $\mathbb{R}^n$. For convenience set $r=\sinh\chi$, $\chi \geq 0$. Then
\begin{align}
 X^a &= \sinh\chi\, \sigma^a(\theta^i) ,  \qquad a = 1,...,n-1\, ,
\\
X^n &= \alpha \cosh\chi \cosh(\tau/\alpha)\, ,
\\
S &=   \alpha \cosh\chi \sinh(\tau/\alpha)\, ,
\\
T &= \sqrt{1+\alpha^2} \cosh\chi\, ,
\end{align}
can be verified to extend to a global embedding ($\chi =0$ is purely a coordinate singularity - the image is a smooth submanifold at $X^a =0$). The pullback of the Minkowski metric w.r.t. this embedding is
\begin{equation}\label{AdSn}
 ds^2 = -\cosh^2 \chi d\tau^2 + d\chi^2 + \sinh^2\chi  d\Omega^2_{n-2}\, ,
\end{equation}
$d\Omega_{n-2}^2$ being the metric of the unit sphere. This is the metric of AdS$_n$.$\qed$

Finally we note that it follows from Lemma \ref{AdSn_lemma}  that there is a one-parameter family of global isometric embeddings of  AdS$_n$ into $\widetilde{\text{AdS}}_{n+1}$.
\\

\textbf{Acknowledgements:} The research leading to these results
has received funding from the European Union Seventh Framework
Programme (FP7/2007-2013) under grant agreement
PCOFUND-GA-2009-246542 and from the Foundation for Science and
Technology of Portugal.

\bibliography{SWP}

\end{document}